%% file: 2019_DK40_Cooperation.tex
\documentclass[a4paper,twoside]{article}

\usepackage{epsfig}
\usepackage{subcaption}
\usepackage{calc}
\usepackage{amssymb}
\usepackage{amstext}
\usepackage{amsmath}
\usepackage{amsthm}
\usepackage{multicol}
\usepackage{multirow}
\usepackage{pslatex}
\usepackage{apalike}
\usepackage{todonotes}
\usepackage{url}
\usepackage{tikz}
\usepackage{pgfplots}
\usepackage[autostyle=true,english=american]{csquotes}
\usepackage[hidelinks]{hyperref}

\usepackage{SCITEPRESS}

\input{setup/settings}

\input{setup/macros}

\begin{document}

\title{Cooperative Maneuvers of Highly Automated Vehicles at Urban Intersections: A Game-theoretic Approach}

%\author{\authorname{XXXX XXXX\sup{1}, XXXX XXXX\sup{1}, XXXX XXXX\sup{1} and XXXX XXXX\sup{1}}
%\affiliation{\sup{1}XXXX, XXXX, XXXX, XXXX}
%\email{\{xxxx,xxxx,xxxx,xxxx\}@xxxx.xxxx}
%}

\author{\authorname{Björn Koopmann, Stefan Puch, Günter Ehmen and Martin Fränzle}
\affiliation{OFFIS e.V., Escherweg 2, 26121 Oldenburg, Germany}
\email{\{koopmann,puch,ehmen,fraenzle\}@offis.de}
}


\abstract{In this paper, we propose an approach how connected and highly automated vehicles can perform cooperative maneuvers such as lane changes and left-turns at urban intersections where they have to deal with human-operated vehicles and vulnerable road users such as cyclists and pedestrians in so-called mixed traffic. In order to support cooperative maneuvers the urban intersection is equipped with an intelligent controller which has access to different sensors along the intersection to detect and predict the behavior of the traffic participants involved. Since the intersection controller cannot directly control all road users and -- not least due to the legal situation -- driving decisions must always be made by the vehicle controller itself, we focus on a decentralized control paradigm. In this context, connected and highly automated vehicles use some carefully selected game theory concepts to make the best possible and clear decisions about cooperative maneuvers. The aim is to improve traffic efficiency while maintaining road safety at the same time. Our first results obtained with a prototypical implementation of the approach in a traffic simulation are promising.}

\onecolumn \maketitle \normalsize \setcounter{footnote}{0} \vfill

\input{sections/sect_introduction}

\input{sections/sect_state_of_the_art}

\input{sections/sect_game_theory}

\input{sections/sect_top-level_architecture}

\input{sections/sect_cooperation_approach}

\input{sections/sect_evaluation}

\input{sections/sect_conclusion}

%\section*{\uppercase{Acknowledgments}}
%
%\noindent xxxx xxxx xxx xxxx xxxxxx xx xxx xxxxxxx xxxxxxxx xx xxxxxxxxx xxx xxxxxxx xxxxxxxxxxxxxx xxxx xx xxxx xx xxx xxxxxxxx xxxxxxxxxxx xxxxxxx xx xxx xxxxxxxxx.
%
%xx xxxxx xxxx xx xxxxx xxx xxxxxxx xxxxxxxx xxx xxxxx xxxxx xxx xxxxxxxxxxx xx xxxxxxxxxxx xx xxx xxxxxxx xxxxxxxxxxx xxxxxxx xxx xxx xxxxx xxxxxxxxx xx xxx xxxx xxxxxx.

\section*{\uppercase{Acknowledgments}}

\noindent This work has been funded by the \emph{Federal Ministry of Transport and Digital Infrastructure} (BMVI) as part of \emph{Digitaler Knoten 4.0} (reference no. 16AVF1008F) and \emph{ViVre} (reference no. 01MM19014E).

We would like to thank all project partners for their trust and cooperation in discussions on the initial cooperation concept and the joint development of the presented reference architecture.

\bibliographystyle{apalike}
{\small
\bibliography{2019_DK40_Cooperation}}

\end{document}

%% file: setup/settings.tex
\pgfplotsset{compat=1.12,
	every axis x label/.style={at={(current axis.right of origin)},anchor=north west},
	every axis y label/.style={at={(current axis.above origin)},anchor=north east}
}

\usetikzlibrary{shapes.geometric}
\usetikzlibrary{tikzmark}

\usepgfplotslibrary{fillbetween}
\usepgfplotslibrary{statistics}

\newcommand{\secref}[1]{Section~\ref{#1}}

\newcommand{\figref}[1]{Figure~\ref{#1}}
\newcommand{\figsref}[2]{Figures~\ref{#1} and~\ref{#2}}
\newcommand{\tabref}[1]{Table~\ref{#1}}

%% file: setup/macros.tex
\newcommand{\eg}{{{e.g.\/,}}}

\newcommand{\AIM}{AIM}
\newcommand{\CHAV}{CHAV}
\newcommand{\CHAVC}{CHAV\nobreakdash-C}
\newcommand{\CHAVs}{CHAVs}
\newcommand{\CLC}{CLC}
\newcommand{\CLCs}{CLCs}

\newcommand{\HV}{HV}
\newcommand{\HVC}{HV\nobreakdash-C}
\newcommand{\HVs}{HVs}
\newcommand{\ITV}{I2V}
\newcommand{\SUMO}{SUMO}
\newcommand{\TMS}{TMS}

\newcommand{\TraCI}{TraCI}

\newcommand{\VRUs}{VRUs}
\newcommand{\VTX}{V2X}

%% file: sections/sect_introduction.tex
\section{\uppercase{Introduction}}
\label{sec:introduction}

\noindent Mobility is a vital basis for individual freedom, an indicator of social prosperity, and an important factor for economic growth \cite{Lemmer2019}. The rising need for mobility of people and goods poses major challenges to administrations of cities and municipalities and stresses the existing infrastructure. In the long term, the structural maintenance and replacement of transport infrastructure is expected to involve increasingly higher investment costs \cite{BMVI2016}.

Besides the growing challenges for municipal administrations, the increased traffic load also reduces the quality of life of the citizens through increased emissions and delays due to congestion. In 2018, German drivers spent on average of more than $120$ hours in traffic jams \cite{INRIX}. At the same time, long congestion periods and unadapted driving styles exacerbate the problems associated with increased air pollution and environmental impact. Discussions on road closures and driving bans resulting from high levels of nitrogen oxides and particulates -- whether effective or not -- can be heard in the press almost every day \cite{Zeit2018,Tagesspiegel2019,Spiegel2019}.

\begin{figure}
	\centering
	\includegraphics[width=0.98\linewidth]{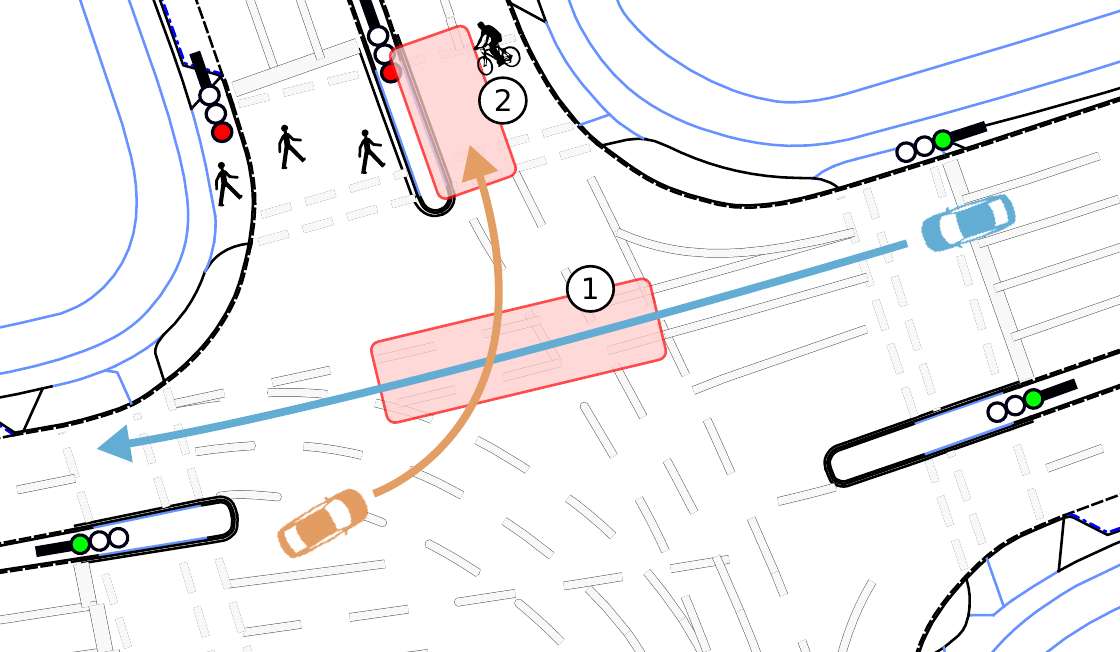}
	\caption{Exemplary Conflict Areas}
	\label{fig:exemplary_conflict_areas}
\end{figure}

Intersections are a key element of today's transport infrastructure and have a significant impact on inner-city traffic efficiency. At these junctions traffic flows from different directions meet, which themselves consist of different road users -- each with individual goals and their own driving styles. Depending on the active traffic light phase, conflict areas with other road users arise while crossing the intersection. \figref{fig:exemplary_conflict_areas} shows an example of these areas resulting from a left-turn scenario. After passing through oncoming traffic (1), the orange-colored vehicle must take crossing cyclists and pedestrians (2) into account.

While human drivers can have difficulties assessing the speeds of other vehicles and the remaining gaps in flowing traffic \cite{YanEtAl2007}, dedicated sensor systems can reliably perform this task. The use of \textit{connected and highly automated vehicles} (\CHAVs{}) in inner-city traffic could therefore help to improve the use of remaining free spaces and to achieve a higher capacity. In addition, however, further challenges arise in the reliable detection of cyclists and pedestrians -- also called \emph{vulnerable road users} (\VRUs{}) -- that share green light phases with motorized road users. Due to the possibility of being covered by vehicles parked at the roadside and the high complexity of their dynamic behavior, this task poses a high degree of difficulty for human drivers in \textit{human-operated vehicles} (\HVs{}) as well as for assistance systems and automated driving functions.

In order to master the challenges of connected and highly automated driving at urban intersections, a common approach is to support the vehicle sensors with infrastructural sensors and communicating information systems to increase sensing ranges and reliability \cite{Burgstrahler2017}. At the same time, this solution enables connected participants to receive comprehensive real-time information about the prevailing traffic situation in remote and poorly visible areas. In addition, traffic light phases as well as information about the future signal course can be transmitted. Equipped with \emph{vehicle-to-everything} (\VTX{}) transceivers and suitable environmental sensors, specialized systems like \emph{traffic management systems} (\TMS{}) could ultimately generate \emph{behavior recommendations} to actively support \CHAVs{} in a safe and efficient crossing of intersection areas.

%While the provision of environmental information through an intelligent and connected traffic infrastructure can increase efficiency, this could undoubtedly be accompanied by higher levels of road safety. Approaching vehicles, for example, can be informed earlier on possible hazards, receive a more reliable situational overview, and can assess the consequences as well as mutual effects of planned driving maneuvers more precisely \cite{ChenEnglund2014}. In this context, the enhanced situational awareness of \CHAVs{} can directly lead to a reduction of the number of critical situations, near misses, and accidents, or mitigate the effects of unavoidable collisions.

Current research gives reason to assume that cooperation of \CHAVs{} with each other, \HVs{}, and \VRUs{} can increase traffic efficiency while ensuring at least the same level of road safety. For this reason, it is likely that in the future not only the automation of individual vehicles, but also their safe interaction with different road users will be an important aspect. With the increasing prevalence of automated vehicles and their growing pervasion in mixed traffic the consideration of cooperation brings new challenges. In this context, the following research questions are of particular importance for the present work:

\begin{enumerate}
	\item How can unambiguous, joint decisions be made whether or not to perform cooperative maneuvers, taking into account global and local goals?
	\item Is it possible to increase traffic efficiency at urban intersections through cooperation?
	\item If so, what conditions must be met in order to promote an efficiency gain? Which conditions may lead to deteriorations?
\end{enumerate}

In order to provide a solution for the first question as well as some hints on answering questions two and three, we will present a novel, decentralized cooperation approach that was developed within the national research project \emph{Digitaler Knoten 4.0} \cite{DigitalerKnoten}. The defined concepts explicitly address the prevalence of mixed traffic that can be expected in the long transition phase of \CHAVs{} to regular operation and could be evaluated directly in the field. We propose a game-theoretic approach which can be implemented algorithmically and allows cooperation candidates to take their own view into account when making decisions about performing a cooperative maneuver or rejecting a specific request. As a prerequisite, we assume the existence of an intelligent \TMS{} equipped with sensors and actuators, which has a global view of the traffic situation and -- based on predictive path planning -- can derive and communicate assessments for meaningful cooperative maneuvers. The approach is evaluated by using a prototypical implementation on the basis of \emph{Simulation of Urban Mobility} (\SUMO{}) \cite{SUMO} that enables us to see some first effects.

This paper is structured as follows. In \secref{sec:state_of_the_art}, existing work on the interaction of \CHAVs{} at urban intersections is discussed. Afterwards, we provide a brief overview of a selected set of game theory concepts. In \secref{sec:top-level_architecture}, the top-level architecture and a description of relevant characteristics of the traffic participants involved are given. \secref{sec:cooperation_approach} presents the developed concepts. This includes the time division of an intersection crossing by a \CHAV{} into phases as well as a detailed investigation of the cooperation approach. In order to analyze the effects of the selected approach, we first describe the implementation of the traffic simulation and the individual experiments in \secref{sec:evaluation}. Subsequently, a discussion of the results is conducted. In \secref{sec:conclusion}, we conclude the paper and give an outlook on future enhancements.

%% file: sections/sect_state_of_the_art.tex
\section{\uppercase{State of the Art}}
\label{sec:state_of_the_art}

\noindent According to \cite{WuEtAl2012}, cooperative driving was first introduced through the use of inter-vehicle communication to perform lane changes and merging maneuvers in the context of \emph{platooning}. It was followed by a line of research to guide vehicles through intersections using the \emph{Autonomous Intersection Management} (\AIM{}) \cite{DresnerStone2008,Dresner2009} approach. The focus is dedicated to avoid critical situations as well as collisions and to adapt the passing sequence of approaching vehicles in order to improve the junction's capacity.

A major challenge regarding the trajectory planning algorithms is noted to be the \enquote{complexity of cooperative driving planning}, because every vehicle has to be considered individually. To this category belong Frese and Beyerer with their work about planning cooperative motions of cognitive automobiles by proposing a tree search algorithm and Grégoire et al. who propose a mathematical framework to decompose the cooperative motion planning problem for vehicles at intersections to \enquote{a discrete scheduling problem (priority graph) and a continuous problem formulated in the abstract coordination space} \cite{FreseBeyerer2010,GregioreEtAl2012}. Kneissl et al. presented a model predictive control based algorithm for automated intersection crossings whereby vehicles do not have to share private data \cite{KneisslEtAl2018}. All these approaches are based on a \emph{central controller} for \CHAVs{} or at least on autonomously executed cooperative maneuvers. Liu et al. extended the previous research by suggesting a hybrid approach where an intersection management system plans collision free trajectories but vehicles are able to arrange their trajectory individually \cite{LiuEtAl2019}. This approach is simplified by the fact that conflict areas first have to be reserved and vehicles drive as platoon over the intersection at constant speed.

None of the previously mentioned approaches considered the execution of cooperative maneuvers in mixed traffic where the somewhat optimal path planning result from \AIM{} has to be combined with uncontrolled traffic participants, whose dynamic behavior is hardly predictable. According to Sharon and Stone, \enquote{\AIM{} has been shown to provide little or no improvement} if less than $90$\,\% of the vehicles are driving autonomously so that the approach cannot be smoothly applied to mixed traffic. They addressed this gap with the development of the \emph{Hybrid Autonomous Intersection Management} protocol \cite{SharonStone2017}. It builds upon a reservation-based \enquote{First Come, First Served} extension of the \AIM{} protocol and shall improve the transition period from mostly \HVs{} to solely \CHAVs{} comprised traffic. In \cite{SharonARBS18}, a centralized manager controls a set of so-called \emph{compliant agents} while other \emph{self-interested agents} coexists within the network. However, cooperative maneuvers between \CHAVs{} are not taken into account.

All existing approaches for the interaction of \CHAVs{} and \HVs{} have in common that they only consider \emph{global goals} such as collision avoidance or an increase of traffic efficiency. They do not take into account the individual pursuit of \emph{local goals}, which may result from manufacturer-specific features or the individual preferences of vehicle occupants. The importance of this limitation is further strengthened by legal and safety constraints of the implementing manufacturers. In this context, a common paradigm is the strict rule that every safety-critical driving decision must be made by the vehicle controller itself. To comply with this demand, all conceivable information provided by other participants and infrastructure systems must only be used to extend a vehicle's \enquote{field of view} and to increase the confidence of its own situation assessment. Only if external requests to achieve global goals are consistent with local goals, then they may influence driving decisions. This style of negotiating cooperative driving maneuvers, taking into account global and local goals while exclusively relying on vehicle-based decision making, is -- to the best of our knowledge -- not subject of current research.

%% file: sections/sect_game_theory.tex
\section{\uppercase{Game Theory}}
\label{sec:game_theory}

\noindent To design a mechanism enabling a joint, unambiguous negotiation process for cooperating \CHAVs{}, we employ a set of commonly used \emph{game theory} concepts \cite{VonNeumannMorgenstern1953}. According to Maschler, game theory subsumes the \enquote{methodology of using mathematical tools to model and analyse situations of interactive decision making} \cite{Maschler2013}. The so-called \emph{players} with possibly different goals influence the further outcome of other players with each of their decision. It can provide valuable services in the analysis of complex economic phenomena as well as in everyday decisions, because it deals with an enormous variety of different decision-making situations and makes suggestions how good decisions can be achieved \cite{Winter2015}.

An important question that has to be answered before applying game theory is which kind of game fits best for the given situation. It is possible to distinguish between \emph{static} or \emph{dynamic} games and games with \emph{complete} or \emph{incomplete} information \cite{Winter2015}. In static games, players act simultaneously without knowing how other players have behaved or will behave. They are also commonly known as \emph{simultaneous} or \emph{concurrent} games \cite{DeAlfaroHenzinger2000}. Conversely, in dynamic games, which are often referred to as \emph{turn-based}, \emph{sequential}, or \emph{repeated} games, players act in a fixed order \cite{DeAlfaroHenzinger2000,Maschler2013}. In games with complete information, each player is well informed about the other players and can practically put himself in their shoes. A player is also aware of all scores, but usually has no knowledge of the individual strategies. Games with incomplete information can result, for example, from strongly restrictive rules, information hiding policies or technical limitations.

Within a \emph{game} each player can have an amount of \emph{strategies} -- good ones which lead to \enquote{win} the game as well as bad ones which lead to \enquote{loose}. A \emph{strategy combination} describes a combination of one strategy per player, a set of all strategy combinations define all possible \emph{game sequences}. Each player can rate his strategies with a measure. This indicates the advantage or benefit of the outcome of the game from the player's perspective. Within the framework of game theory, this measure is referred to as \emph{payoff}.

The objective of game theory is to find an optimal strategy combination based on the payoffs of the individual players. From a player's point of view, the primary goal is to achieve the highest possible payoff by selecting the ``best'' strategy combination in order to ``win'' the game. In this context, an optimal strategy combination from a global perspective is referred to as \emph{Nash equilibrium} \cite{Nash1951}. If the equilibrium is played, none of the players would unilaterally change their mind, because no other strategy could achieve a higher payoff or an improvement. It therefore allows distributed strategy finding without additional communication, just by rational reasoning.

A common representation for games with a limited number of strategies are so-called \emph{payoff matrices}. They list the strategies of one player in rows and the strategies of another in columns. Each cell thus contains the combined payoffs for the combination of both individual strategies. We will use this form of representation and the underlying theory to select the best possible strategy combination and to decide in a distributed manner whether or not to perform the resulting cooperative driving maneuver.

%% file: sections/sect_top-level_architecture.tex
\section{\uppercase{Top-Level Architecture}}
\label{sec:top-level_architecture}

\noindent In order to control the flow of mixed traffic at urban intersections, the \emph{Digitaler Knoten 4.0} project consortium developed a reference architecture, which can be understood as a blueprint for the digitalization of urban intersections. Here, the extensive expert knowledge and know-how from previous projects of the industrial and academic partners involved were used to provide a comprehensive and consistent architecture to guide the development of future transport systems.

\begin{figure}
	\centering
	\includegraphics[width=0.98\linewidth]{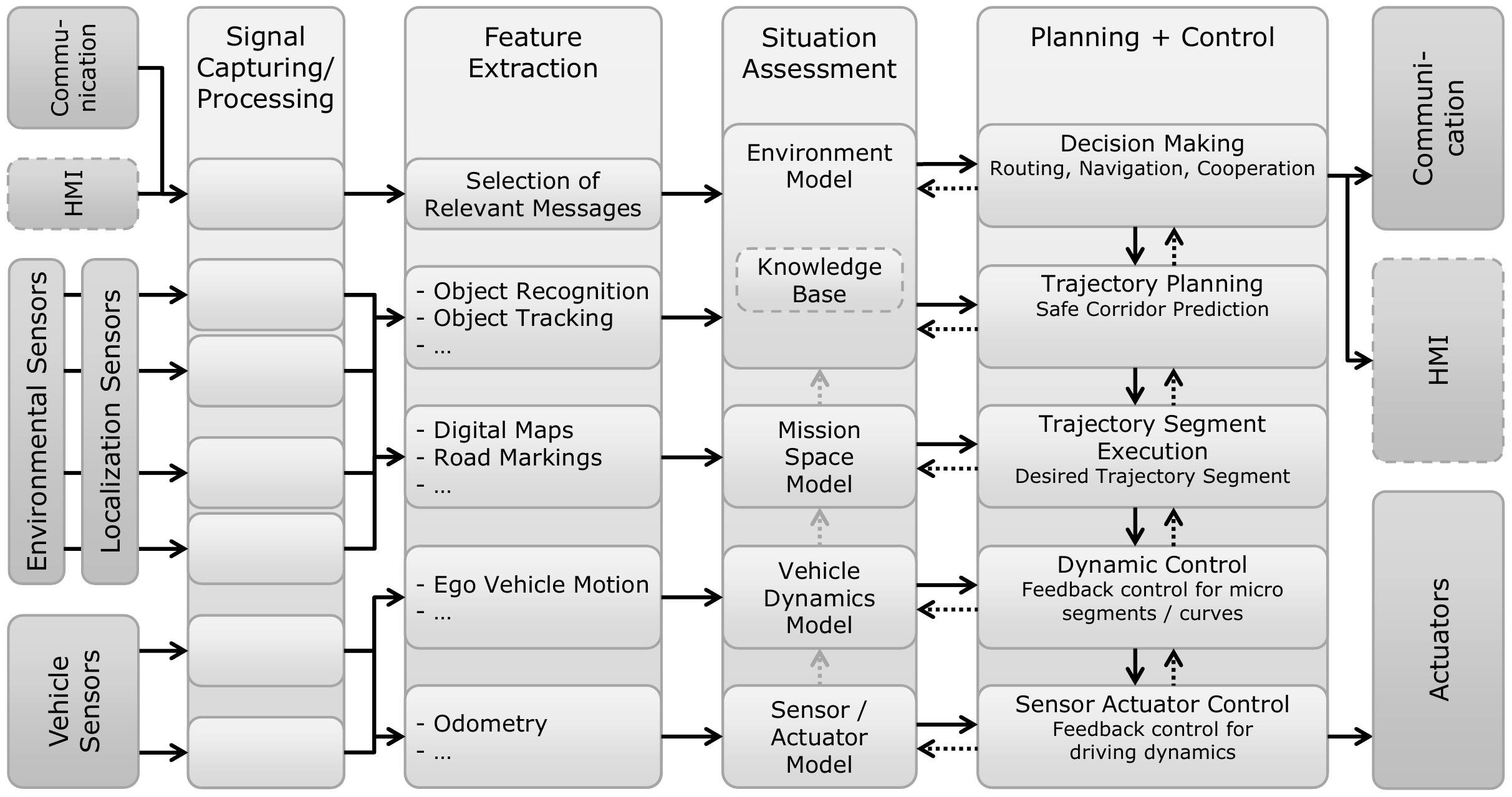}
	\caption{(Sub-)Architecture \emph{Automated Vehicle}}
	\label{fig:sub-architecture_automated_vehicle}
\end{figure}

At the highest level, the proposed reference architecture is divided into two (sub-)architectures \emph{Automated Vehicle} and \emph{Traffic Management System}, which are illustrated in \figsref{fig:sub-architecture_automated_vehicle}{fig:sub-architecture_traffic_management_system}. While the communication between these subsystem classes is realized by means of a third reference architecture \emph{Communication Channel}, their individual components are not explained in detail for the sake of simplicity.

\begin{figure}
	\centering
	\includegraphics[width=0.98\linewidth]{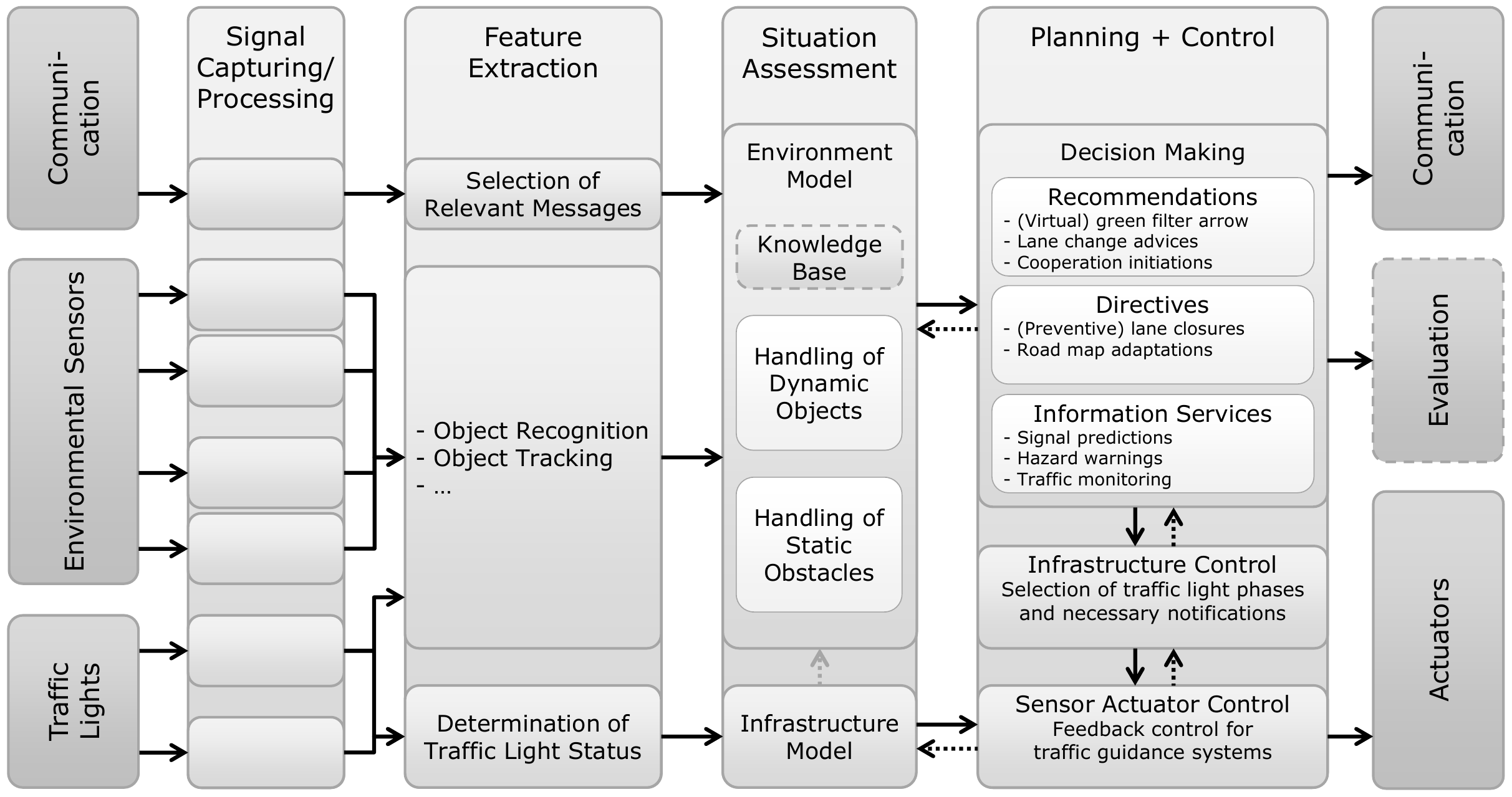}
	\caption{(Sub-)Architecture \emph{Traffic Management System}}
	\label{fig:sub-architecture_traffic_management_system}
\end{figure}

The two architectures in focus are layered architectures whose horizontal design is based on the commonly used \enquote{Sense-Plan-Act} paradigm. In this context, each subarchitecture consists of six columns that structure the underlying components according to their functional purpose. Triggered by the input events of the sensor components, the collected information is propagated successively through the individual segments and is aggregated to an encompassing real-time situational overview on which either driving or control decisions can be made.

The (abstract) functional components, each representing a set of possible implementations of required tasks, are arranged at different abstraction levels according to their time resolution or activation frequency -- ranging from control-based algorithms on lower levels to service-oriented concepts on higher layers. While the first subarchitecture contains up to five abstraction levels, the \emph{Traffic Management System} has a simplified vertical design due to the lack of proper motion and the comparatively low complexity of the control tasks assigned to it.

In the following, some general properties of the considered traffic participants will be put into concrete terms. While the reference architecture already provides a good overview of the functionality of the technical systems involved, the assumptions used about the equipment of the \TMS{} have to be concretized. In addition, suitable assumptions about the physical properties and the dynamic behavior of \CHAVs{}, \HVs{}, and \VRUs{} will be formulated.

\subsection{Traffic Management System}
\label{subsec:traffic_management_system}

As explained above, the use of an intelligent \TMS{} is a key element of our approach. In our vision, such a system has the capability to send and receive \VTX{} messages with a minimum reception range of $200$\,m. In addition, we assume it to be equipped with environmental sensors that detect and track static obstacles and dynamic objects within a sensing range of $150$\,m in order to maintain a real-time situational overview and to predict the future behavior of all participants. As a third source of information, a \TMS{} should be linked to the traffic light system, which provides access to the signal phases and future signal courses.

Based on the situational overview aggregated in the environment and infrastructure models, the \emph{Decision Making} component analyzes the current and anticipated traffic situation in the intersection area. It brings intelligence to the \TMS{} and takes all high-level control decisions related to the flow of mixed traffic. Besides the transmission of \emph{infrastructure-to-vehicle} (\ITV{}) messages, such as \emph{behavior recommendations} and \emph{hazard warnings}, it controls the infrastructure actuators (\eg{} variable message signs), and provides evaluation services for traffic analyses.

\subsection{Highly Automated Vehicles}
\label{subsec:highly_automated_vehicles}

While the \CHAVs{} in focus are assumed to have automation levels from $4$ to $5$ \cite{SAE2018}, their individual driving tasks are (almost) completely taken over. In accordance with the explanations on the \TMS{}, all decisions are drawn by a central \emph{Decision Making} component. Every \CHAV{} has the capability to send and receive \VTX{} messages and comes with a suitable set of environmental sensors. %Moreover, a high-accuracy self-localization as well as state-of-the-art vehicle sensors are considered as prerequisites for its highly-automated operation in inner-city traffic.

Each \CHAV{} moves at a maximum speed of $50$\,km/h and complies with all traffic regulations. Due to the high degree of automation and the lack of human reaction time, a \CHAV{} should also adapt to external events in significantly less than $1$\,s. At the same time, conservative estimates of distances and time gaps are used -- the vehicle thus drives very defensively and with foresight. While all \CHAVs{} behave prudently when interacting with other road users, it is assumed that no cooperative maneuvers are performed without using the proposed approach.

%While all \CHAVs{} behave prudently when interacting with each other and non-automated participants, it is assumed in the following that no cooperative driving maneuvers are performed without using the approach proposed in \secref{sec:cooperation_approach}.

\subsection{Human-operated Vehicles}
\label{subsec:human-operated_vehicles}

In contrast to \CHAVs{}, all \HVs{} are assumed to have automation levels ranging from $0$ to $3$ \cite{SAE2018}. The driving task is therefore primarily taken over by the human driver. For this reason, the driving behavior is significantly influenced by his personality and state of health as well as his local knowledge and individual goals, \eg{} resulting from time pressure or being on a \enquote{sightseeing tour}. The communication with other road users is usually done by simple light and hand signals. Although top-of-the-range vehicles may already have the option to receive \VTX{} or \ITV{} messages at these automation levels, it is not considered to be the case for the majority.

Depending on the passenger's preferences, a typical \HV{} drives at a higher maximum speed of up to $60$\,km/h. In most cases, the driver adheres to the traffic rules, but deviates from them in certain situations. In addition, his reaction time is about $1$\,s. During the journey of a \HV{} the driver under- or overestimates distances and time gaps due to the natural limitations of his perception. The possibility of mental overload in unexpected situations results in a higher failure probability compared to \CHAVs{}.

\subsection{Cyclists and Pedestrians}
\label{subsec:cyclists_and_pedestrians}

The non-motorized road users, consisting of cyclists and pedestrians, exhibit a comparatively low speed compared to \CHAVs{} and \HVs{}. At the same time, the safe prediction of their dynamic behavior poses a major challenge due to the high complexity of the underlying models and the possibility of fast directional changes. While appropriate concepts to handle non-compliant behaviors exist, we limit ourselves to the assumption that \VRUs{} move on bicycle paths or sidewalks and behave according to the traffic rules.

%% file: sections/sect_cooperation_approach.tex
\section{\uppercase{Cooperation Approach}}
\label{sec:cooperation_approach}

\begin{figure}
	\centering
	\includegraphics[width=0.98\linewidth]{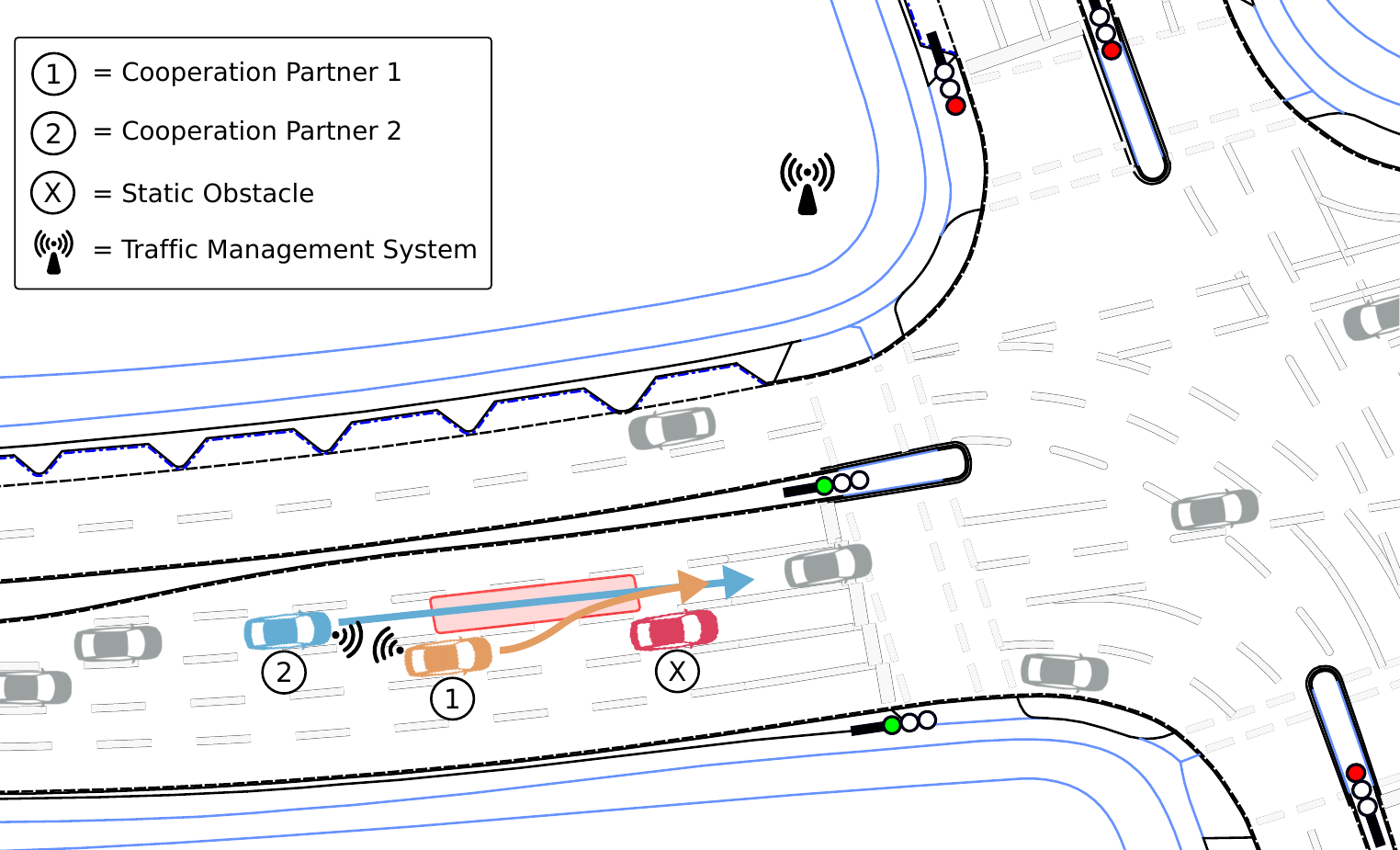}
	\caption{Cooperative Lane Change}
	\label{fig:cooperative_lane_change}
\end{figure}

\noindent There is a wide variety of conceivable scenarios that can provide benefits for the participating road users and municipal administrations. In the further course, the \emph{cooperative lane change} (\CLC{}), whose basic principle is illustrated in \figref{fig:cooperative_lane_change}, will be subjected to closer examination. It was extensively studied as a further part of the \emph{Digitaler Knoten 4.0} project.

At the beginning of the scenario, the orange-colored \CHAV{} \raisebox{.5pt}{\textcircled{\raisebox{-.9pt} {1}}} approaches a static obstacle \raisebox{.5pt}{\textcircled{\raisebox{-.9pt} {$\times$}}}, such as a vehicle at the end of a congestion or a narrowing of the roadway. While the \CHAV{} in focus would normally have to wait, a negotiation with the turquoise-colored \CHAV{} \raisebox{.5pt}{\textcircled{\raisebox{-.9pt} {2}}} on the adjacent lane allows an agreement on performing a \CLC{} that enables an efficient merge into flowing traffic. In order to avoid unnecessary strong effects on the second cooperation partner or other dangerous situations, a thorough analysis of the traffic situation and suitable concepts for a distributed assessment are required.

While the interaction of multiple traffic participants would be conceivable and could have more far-reaching effects on traffic efficiency, we will focus on the cooperation of only two \CHAVs{} to explain the basic principles of the underlying concepts.

\subsection{Time Subdivision into Phases}
\label{subsec:time_subdivision_into_phases}

In order to guide the development of the overall cooperation approach and to enable a structured analysis of the resulting traffic situations, we decided to divide the journey of each \CHAV{} through the intersection area into three so-called \emph{cooperation phases}.

\subsubsection{Subscription Phase}
\label{subsubsec:subscription_phase}

The aim of this first phase is to improve the situational awareness of the \CHAV{} in focus and the \TMS{} by communicating vehicle-specific data and exchange information about the traffic situation to set the stage for cooperative maneuvers. In a first step, a unique \emph{vehicle ID}, a \emph{global driving intention} (\eg{} turn left, straight ahead, turn right), and the \emph{intended destination lane} are sent to the \TMS{}. The reception of the message is confirmed to the \CHAV{} by sending a detailed response containing the current \emph{signal phase} of the traffic lights, their future \emph{signal course}, a \emph{digital map} of the intersection as well as comprehensive information about \emph{static obstacles} and \emph{dynamic objects}.

According to the assumptions made in \secref{subsec:traffic_management_system}, it is assumed hereinafter that the \emph{subscription phase} is completed at least $150$\,m before the stop line of the respective lane. If the subscription of a \CHAV{} is not completed in time, the crossing must take place without making any benefits from the additional services provided by the \TMS{}, including the support in cooperative maneuvers. Of course, an adaptation of the assumed range constraint is possible to accommodate other intersection geometries.

\subsubsection{Execution Phase}
\label{subsubsec:execution_phase}

From the perspective of a \CHAV{}, the primary aim of the \emph{execution phase} is to cross the intersection safely and as efficiently as possible. During the rule-compliant approach of the \CHAV{} to the stop line, the interaction with other road users takes place as it would do even without the existence of an intelligent \TMS{}. Based on the information acquired in the first phase, however, an earlier decision can be made for or against passing the intersections traffic lights.

At the same time, the \TMS{} aims to increase the junction's capacity, prevent congestion, and reduce emissions. It makes use of all available information about the current positions, velocities, driving intentions, and the planned destination lanes of the subscribed \CHAVs{} and other participants, its knowledge about the intersection geometry, static obstacles as well as dynamic objects, and identifies potential conflict situations between \CHAVs{} to be solved. In a second step, the \TMS{} tries to derive feasible solutions to these conflicts that increases traffic efficiency, \eg{} through behavioral predictions and the recognition of previously learned patterns within the traffic scene.

Subsequently, promising solutions are delivered to the participating \CHAVs{} in the form of \emph{cooperation recommendations}. The \CHAVs{} initiate distributed assessments of the traffic situations based on their own situational overviews and exchange their \emph{evaluations}, which can ultimately lead to joint decisions and the execution of the proposed cooperative maneuvers.

\subsubsection{Unsubscription Phase}
\label{subsubsec:unsubscription_phase}

In the course of the \emph{unsubscription phase}, the information collected during the crossing of the intersection is fed back to the \TMS{}. By receiving the periodically sent positions and velocities of all exiting \CHAVs{}, further information can be obtained on traffic density and the likelihood of congestion in the individual exits. After leaving the reception area of the \VTX{} transceiver, each \CHAV{} is indirectly unsubscribed by its removal from the situational overview.

\subsection{Cooperation Procedure}
\label{subsec:cooperation_procedure}

In contrast to almost all existing publications, we propose a decentralized cooperation approach for the negotiation of cooperative maneuvers. For this purpose, the \TMS{} serves as a central source of information and recommendations that actively supports the approaching \CHAVs{} as they cross the intersection. At the same time, it includes other traffic participants using suitable sensors and actuators. The involved \CHAVs{} agree decentrally on recommended driving maneuvers or decide against their execution.

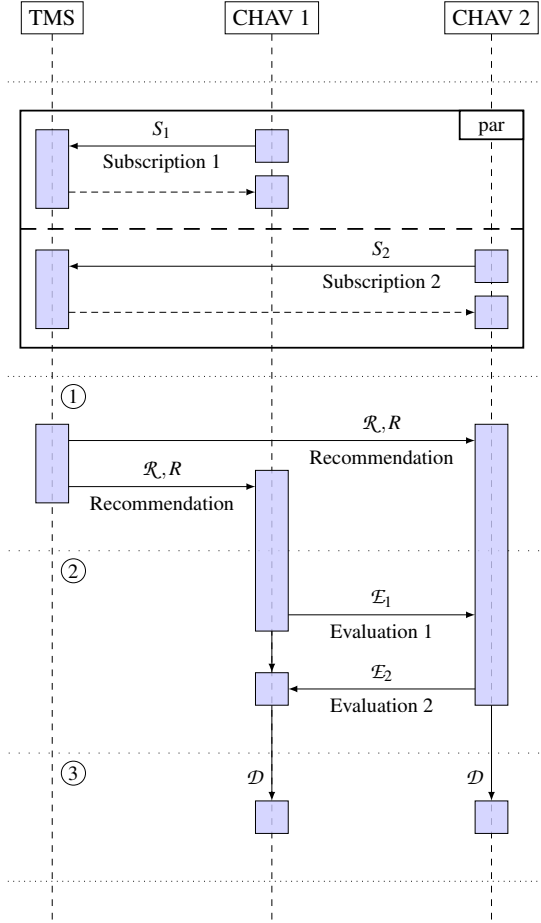
\begin{figure}
	\centering
	
	\def\TMS{TMS}
	\def\CHAVone{CHAV1}
	\def\CHAVtwo{CHAV2}
	
	\def\Subscription{Subscription}
	\def\Recommendation{Recommendation}
	\def\Evaluation{Evaluation}
	
	\begin{tikzpicture}[every node/.style={font=\normalsize,minimum height=0.5cm,minimum width=0.5cm}]
		\clip (-3.5,-5.88) rectangle (3.52,6.4);
		
		\begin{scope}[scale=0.85,transform canvas={scale=0.85}]
			\node [matrix, very thin,column sep=1.2cm,row sep=0.5cm] (matrix) at (0,0) {
				% +++ LIFELINE DESCRIPTIONS +++
				& \node(0,0) (\TMS) {}; & & \node(0,0) (\CHAVone) {}; & & \node(0,0) (\CHAVtwo) {}; & \\

				% +++ SUBSCRIPTION PHASE +++
				\node(0,0) (t0 left) {}; & & & & & & \node(0,0) (t0 right) {};\\
				
					% Subscription request of CHAV 1
					& \node(0,0) (\TMS S1) {}; & \node(0,0) (\Subscription 1) {}; & \node(0,0) (\CHAVone S1) {}; & & \node(0,0) (\CHAVtwo S1) {}; & \\[-8pt]
					
					% Subscription acknowledgment for CHAV 1
					& \node(0,0) (\TMS S2) {}; & & \node(0,0) (\CHAVone S2) {}; & & \node(0,0) (\CHAVtwo S2) {}; & \\[-8pt]
				
				\node(0,0) (tS1 left) {}; & & & & & & \node(0,0) (tS1 right) {};\\[-16pt]
				
					% Subscription request of CHAV 2
					& \node(0,0) (\TMS S3) {}; & & \node(0,0) (\CHAVone S3) {}; & \node(0,0) (\Subscription 2) {}; & \node(0,0) (\CHAVtwo S3) {}; & \\[-8pt]
					
					% Subscription acknowledgment for CHAV 2
					& \node(0,0) (\TMS S4) {}; & & \node(0,0) (\CHAVone S4) {}; & & \node(0,0) (\CHAVtwo S4) {}; & \\
				
				% +++ EXECUTION PHASE +++
				\node(0,0) (tS3 left) {}; & & & & & & \node(0,0) (tS3 right) {};\\
				
					% Recommendation for CHAV 2
					& \node(0,0) (\TMS E1) {}; & & \node(0,0) (\CHAVone E1) {}; & \node(0,0) (\Recommendation 2) {}; & \node(0,0) (\CHAVtwo E1) {}; & \\[-8pt]
					
					% Recommendation for CHAV 1
					& \node(0,0) (\TMS E2) {}; & \node(0,0) (\Recommendation 1) {}; & \node(0,0) (\CHAVone E2) {}; & & \node(0,0) (\CHAVtwo E2) {}; & \\
				
				\node(0,0) (tE1 left) {}; & & & & & & \node(0,0) (tE1 right) {};\\
				
					% Evaluation of CHAV 1
					& \node(0,0) (\TMS E3) {}; & & \node(0,0) (\CHAVone E3) {}; & \node(0,0) (\Evaluation 1) {}; & \node(0,0) (\CHAVtwo E3) {}; & \\[-8pt]
					
					\node(0,0) (tE2 left) {}; & & & & & & \node(0,0) (tE2 right) {};\\[-16pt]
					
					% Evaluation of CHAV 2
					& \node(0,0) (\TMS E4) {}; & & \node(0,0) (\CHAVone E4) {}; & \node(0,0) (\Evaluation 2) {}; & \node(0,0) (\CHAVtwo E4) {}; & \\
				
				\node(0,0) (tE4 left) {}; & & & & & & \node(0,0) (tE4 right) {};\\
				
					% Execution of cooperative maneuver
					& \node(0,0) (\TMS E5) {}; & & \node(0,0) (\CHAVone E5) {}; & & \node(0,0) (\CHAVtwo E5) {}; & \\
				
				\node(0,0) (tE5 left) {}; & & & & & & \node(0,0) (tE5 right) {};\\

				% +++ LIFELINE ENDINGS +++
				& \node(0,0) (\TMS E6) {}; & & \node(0,0) (\CHAVone E6) {}; & & \node(0,0) (\CHAVtwo E6) {}; & \\
			};
			
			\fill (\TMS) node[draw, fill=white] {\TMS{}};
			\fill (\CHAVone) node[draw, fill=white] {\CHAV{} 1};
			\fill (\CHAVtwo) node[draw, fill=white] {\CHAV{} 2};
			
			% Horizontal time lines
			\draw [dotted] (t0 left) -- (t0 right) node[right] {};
			\draw [dotted] (tS3 left) -- (tS3 right) node[right] {};
			\draw [loosely dotted] (tE1 left) -- (tE1 right) node[right] {};
			\draw [loosely dotted] (tE4 left) -- (tE4 right) node[right] {};
			\draw [dotted] (tE5 left) -- (tE5 right) node[right] {};
			
			% Vertical life lines
			\draw [dashed] (\TMS) -- (\TMS E6);
			\draw [dashed] (\CHAVone) -- (\CHAVone E6);
			\draw [dashed] (\CHAVtwo) -- (\CHAVtwo E6);
			
			% Vertical flows
			\draw [-latex] (\CHAVone E3) -- (\CHAVone E4);
			\draw [-latex] (\CHAVone E4) -- (\CHAVone E5) node[pos=0.75, left]{\small $\mathcal{D}$};
			\draw [-latex] (\CHAVtwo E4) -- (\CHAVtwo E5) node[pos=0.75, left]{\small $\mathcal{D}$};
			
			% Blocks (subscription phase)
			\filldraw [fill=blue!20, fill opacity=0.8] (\TMS S1.north west) rectangle (\TMS S2.south east);
			\filldraw [fill=blue!20, fill opacity=0.8] (\TMS S3.north west) rectangle (\TMS S4.south east);
			
			\filldraw [fill=blue!20, fill opacity=0.8] (\CHAVone S1.north west) rectangle (\CHAVone S1.south east);
			\filldraw [fill=blue!20, fill opacity=0.8] (\CHAVone S2.north west) rectangle (\CHAVone S2.south east);
			
			\filldraw [fill=blue!20, fill opacity=0.8] (\CHAVtwo S3.north west) rectangle (\CHAVtwo S3.south east);
			\filldraw [fill=blue!20, fill opacity=0.8] (\CHAVtwo S4.north west) rectangle (\CHAVtwo S4.south east);
			
			% Blocks (execution phase)
			\filldraw [fill=blue!20, fill opacity=0.8] (\TMS E1.north west) rectangle (\TMS E2.south east);
			
			\filldraw [fill=blue!20, fill opacity=0.8] (\CHAVone E2.north west) rectangle (\CHAVone E3.south east);
			\filldraw [fill=blue!20, fill opacity=0.8] (\CHAVone E4.north west) rectangle (\CHAVone E4.south east);
			\filldraw [fill=blue!20, fill opacity=0.8] (\CHAVone E5.north west) rectangle (\CHAVone E5.south east);
			
			\filldraw [fill=blue!20, fill opacity=0.8] (\CHAVtwo E1.north west) rectangle (\CHAVtwo E4.south east);
			\filldraw [fill=blue!20, fill opacity=0.8] (\CHAVtwo E5.north west) rectangle (\CHAVtwo E5.south east);
			
			% Horizontal flows (subscription phase)
			\draw [-latex] (\CHAVone S1) -- (\TMS S1);
			\draw [-latex, densely dashed] (\TMS S2) -- (\CHAVone S2);
			
			\draw [-latex] (\CHAVtwo S3) -- (\TMS S3);
			\draw [-latex, densely dashed] (\TMS S4) -- (\CHAVtwo S4);
			
			% Horizontal flows (execution phase)
			\draw [-latex] (\TMS E1) -- (\CHAVtwo E1);
			\draw [-latex] (\TMS E2) -- (\CHAVone E2);
			
			\draw [-latex] (\CHAVone E3) -- (\CHAVtwo E3);
			
			\draw [-latex] (\CHAVtwo E4) -- (\CHAVone E4);
			
			% Flows labels (subscription phase)
			\fill (\Subscription 1) node[font=\small, above] {$S_1$} node[font=\small, below] {\Subscription~1};
			\fill (\Subscription 2) node[font=\small, above] {$S_2$} node[font=\small, below] {\Subscription~2};
			
			% Flows labels (execution phase)
			\fill (\Recommendation 1) node[font=\small, above] {$\mathcal{R},R$} node[font=\small, below] {\Recommendation};
			\fill (\Recommendation 2) node[font=\small, above] {$\mathcal{R},R$} node[font=\small, below] {\Recommendation};
			
			\fill (\Evaluation 1) node[font=\small, above] {$\mathcal{E}_1$} node[font=\small, below] {\Evaluation~1};
			
			\fill (\Evaluation 2) node[font=\small, above] {$\mathcal{E}_2$} node[font=\small, below] {\Evaluation~2};
			
			% Parallel combined fragment
			\draw [thick] ([xshift=-8, yshift=10] \TMS S1.north west) rectangle ([xshift=8, yshift=-10] \CHAVtwo S4.south east);
			\draw [thick, dash pattern=on 7pt off 5pt] ([xshift=-8, yshift=11] \TMS S3.north west) -- ([xshift=8, yshift=11] \CHAVtwo S3.north east);
			\draw [thick, fill=white] ([xshift=-8, yshift=10] \CHAVtwo S1.north west) rectangle ([xshift=8, yshift=12] \CHAVtwo S1.south east) node[pos=0.5, yshift=-1.5] {\small par};
			
			% Step numbering
			\draw ([xshift=2.5, yshift=14.5] \TMS E1.north east) circle [thick, radius=0.22] node {\small $1$};
			\draw ([xshift=2.5, yshift=14.5] \TMS E3.north east) circle [thick, radius=0.22] node {\small $2$};
			\draw ([xshift=2.5, yshift=14.5] \TMS E5.north east) circle [thick, radius=0.22] node {\small $3$};
		\end{scope}
	\end{tikzpicture}
	\caption{Time Sequence of Communication}
	\label{fig:time_sequence_of_communication}
\end{figure}

In order to give a clearer insight into the underlying concepts, \figref{fig:time_sequence_of_communication} provides an overview of the intended communication flow. Following the distribution of a \emph{cooperation recommendation} to the two involved \CHAVs{} \raisebox{.5pt}{\textcircled{\raisebox{-.9pt} {1}}}, an assessment of the proposed cooperative maneuver and the resulting strategy combinations is performed locally. Each vehicle then communicates the results of its assessment by means of an \emph{evaluation} \raisebox{.5pt}{\textcircled{\raisebox{-.9pt} {2}}}. The messages sent are also received by the \TMS{} and can be used to monitor the cooperation process and improve behavioral predictions. As a final step, the decision is made for or against the execution of the cooperative maneuver \raisebox{.5pt}{\textcircled{\raisebox{-.9pt} {3}}}.

\subsubsection{Cooperation Recommendation}
\label{subsubsec:cooperation_recommendation}

As already described in \secref{subsubsec:execution_phase}, the \TMS{} makes use of all available information about the current positions, velocities, driving intentions, and the planned destination lanes of traffic participants, its knowledge about the intersection geometry, static obstacles as well as dynamic objects, and identifies potential conflict situations between \CHAVs{} to be solved. If such a situation is identified in which an increase in efficiency can be expected through the execution of a cooperative driving maneuver, it is first checked whether the vehicles involved are \CHAVs{}. If so, their unique IDs are stored to prepare the message exchange. Additionally, the vehicle benefiting directly from increased traffic efficiency is designated as \emph{first cooperation candidate} (\CHAV{} 1), while the other one is denoted as \emph{second cooperation candidate} (\CHAV{} 2).

In the further course, suitable strategies for both vehicles are identified or extracted from a strategy catalogue based on expert knowledge and feedback received from preceding vehicles during their individual unsubscription phases. A strategy consists of a pair of (abstract) longitudinal and lateral behavior descriptions, which can be derived directly from the cooperative maneuver and the given traffic situation.

With regard to the example shown in \figref{fig:cooperative_lane_change}, the vehicle approaching the obstacle, which is referred to as \CHAV{} 1, is recommended to perform a \CLC{} with \CHAV{} 2. \tabref{tab:strategies_for_chav_1} contains a list of exemplary strategies that can be used to solve the imminent conflict situation from the perspective of \CHAV{} 1.

\begin{table}[h]
	%\vspace{-0.2cm}
	\caption{Strategies for \CHAV{} 1}
	\label{tab:strategies_for_chav_1}
	\centering
	\begin{tabular}{|c|p{1.8cm}|p{1.8cm}|}
		\hline
		ID & Longitudinal \newline Behavior & Lateral \newline Behavior \\
		\hline
		S1.1 & \textit{Continue} & \textit{Lane change} \\
		\hline
		S1.2 & \textit{Decelerate} & \textit{Continue} \\
		\hline
	\end{tabular}
\end{table}

On the other hand, the vehicle on the adjacent lane, denoted as \CHAV{} 2, is recommended to support the \CLC{} of its cooperation partner \CHAV{} 1. Again, a selected set of suitable strategies is listed in \tabref{tab:strategies_for_chav_2}. Of course, in realistic situations, considerably more strategies would have to be taken into account, such as changing lanes with simultaneous accelerations or varying degrees of intensity of certain maneuvers.

\begin{table}[h]
	%\vspace{-0.2cm}
	\caption{Strategies for \CHAV{} 2}
	\label{tab:strategies_for_chav_2}
	\centering
	\begin{tabular}{|c|p{1.8cm}|p{1.8cm}|}
		\hline
		ID & Longitudinal \newline Behavior & Lateral \newline Behavior \\
		\hline
		S2.1 & \textit{Decelerate} & \textit{Continue} \\
		\hline
		S2.2 & \textit{Continue} & \textit{Continue} \\
		\hline
	\end{tabular}
\end{table}

Following the derivation of suitable strategies for both participants, the initial payoff matrix $\mathcal{R}$ is generated, which is shown in \tabref{tab:initial_payoff_matrix}. Each of its cells represents a strategy combination that results from the selection of one strategy for each cooperation candidate. At this point of time, all entries are empty, since no valuations of the strategy combinations exist.

\begin{table}[h]
	%\vspace{-0.2cm}
	\caption{Initial Payoff Matrix}
	\label{tab:initial_payoff_matrix}
	\centering
	\begin{tabular}{|c|c|c|c|}
		\hline
		
		\multicolumn{2}{|c|}{\multirow{2}{*}[-3.2pt]{$\mathcal{R}$}} & \multicolumn{2}{|c|}{} \\[-9pt]
		
		\multicolumn{2}{|c|}{} & \multicolumn{2}{|c|}{\CHAV{} 2} \\
		
		\cline{3-4}
		
		\multicolumn{2}{|c|}{} & & \\[-9pt]
		
		\multicolumn{2}{|c|}{} & S2.1 & S2.2 \\
		
		\hline
		
		\multirow{2}{*}{\rotatebox[origin=c]{90}{\CHAV{} 1}} &
		\rotatebox[origin=c]{90}{~S1.1~} &
		$(\text{~--~}, \text{~--~})$ &
		$(\text{~--~}, \text{~--~})$ \\
		
		\cline{2-4}
		
		& \rotatebox[origin=c]{90}{~S1.2~} &
		$(\text{~--~}, \text{~--~})$ &
		$(\text{~--~}, \text{~--~})$ \\
		
		\hline
	\end{tabular}
\end{table}

As a final step, the initial payoff matrix $\mathcal{R}$ as well as the recommended strategy combination $R$, which represents the most valuable solution in terms of global goals, are transmitted to the \CHAVs{}.

\subsubsection{Evaluation and Decision Making}
\label{subsubsec:evaluation_and_decision_making}

After receiving the recommendation, an independent, in-vehicle evaluation of all strategy combinations generated by the \TMS{} is initiated. For this purpose, each cooperation candidate evaluates the solutions based of its own situational overview. Besides the speeds of the road users involved, distances between them and differential speeds, the predicted future behavior of dynamic objects and local goals of the \CHAV{} could be included. While the evaluation approach is manufacturer- and implementation-specific, the payoffs can be understood as a function of the results of two separate safety and efficiency evaluations. The calculated payoffs are then filled into the matrix entries as illustrated in \tabref{tab:evaluation_of_the_cooperative_maneuver}.

\begin{table}[h]
	%\vspace{-0.2cm}
	\caption{Evaluation of the Cooperative Maneuver}
	\label{tab:evaluation_of_the_cooperative_maneuver}
	\centering
	\begin{tabular}{|c|c|c|c|}
		\hline
		
		\multicolumn{2}{|c|}{\multirow{2}{*}[-3.2pt]{$\mathcal{E}_1$}} & \multicolumn{2}{|c|}{} \\[-9pt]
		
		\multicolumn{2}{|c|}{} & \multicolumn{2}{|c|}{\CHAV{} 2} \\
		
		\cline{3-4}
		
		\multicolumn{2}{|c|}{} & & \\[-9pt]
		
		\multicolumn{2}{|c|}{} & S2.1 & S2.2 \\
		
		\hline
		
		\multirow{2}{*}{\rotatebox[origin=c]{90}{\CHAV{} 1}} &
		\rotatebox[origin=c]{90}{~S1.1~} &
		$(\text{~4~}, \text{~--~})$ &
		$(\text{~3~}, \text{~--~})$ \\
		
		\cline{2-4}
		
		& \rotatebox[origin=c]{90}{~S1.2~} &
		$(\text{-2~}, \text{~--~})$ &
		$(\text{~1~}, \text{~--~})$ \\
		
		\hline
	\end{tabular}
	
	\vspace{4pt}
	
	\begin{tabular}{|c|c|c|c|}
		\hline
		
		\multicolumn{2}{|c|}{\multirow{2}{*}[-3.2pt]{$\mathcal{E}_2$}} & \multicolumn{2}{|c|}{} \\[-9pt]
		
		\multicolumn{2}{|c|}{} & \multicolumn{2}{|c|}{\CHAV{} 2} \\
		
		\cline{3-4}
		
		\multicolumn{2}{|c|}{} & & \\[-9pt]
		
		\multicolumn{2}{|c|}{} & S2.1 & S2.2 \\
		
		\hline
		
		\multirow{2}{*}{\rotatebox[origin=c]{90}{\CHAV{} 1}} &
		\rotatebox[origin=c]{90}{~S1.1~} &
		$(\text{~--~}, \text{~2~})$ &
		$(\text{~--~}, \text{-2~})$ \\
		
		\cline{2-4}
		
		& \rotatebox[origin=c]{90}{~S1.2~} &
		$(\text{~--~}, \text{~1~})$ &
		$(\text{~--~}, \text{~0~})$ \\
		
		\hline
	\end{tabular}
\end{table}

Once all strategy combinations have been successfully assessed, the exchange of the (so far only half-filled) payoff matrices $\mathcal{E}_1$ and $\mathcal{E}_2$ proceeds in order to merge them into two complete and consistent copies of matrix $\mathcal{D}$, one for each vehicle. In order to make the game fair and safe, communication must take place (almost) simultaneously. For the practical implementation of this approach, it is crucial that no prematurely received evaluations of other players must be used as a basis for one's own evaluation. In terms of game theory, it has to be a \emph{static} game.

\begin{table}[h]
	%\vspace{-0.2cm}
	\caption{Decision Making}
	\label{tab:decision_making}
	\centering
	\begin{tabular}{|c|c|c|c|}
		\hline
		
		\multicolumn{2}{|c|}{\multirow{2}{*}[-3.2pt]{$\mathcal{D}$}} & \multicolumn{2}{|c|}{} \\[-9pt]
		
		\multicolumn{2}{|c|}{} & \multicolumn{2}{|c|}{\CHAV{} 2} \\
		
		\cline{3-4}
		
		\multicolumn{2}{|c|}{} & & \\[-9pt]
		
		\multicolumn{2}{|c|}{} & S2.1 & S2.2 \\
		
		\hline
		
		\multirow{2}{*}{\rotatebox[origin=c]{90}{\CHAV{} 1}} &
		\rotatebox[origin=c]{90}{~S1.1~} &
		\tikzmark{startup}$(\text{~4~}, \text{~2~})$\tikzmark{endup} &
		$(\text{~3~}, \text{-2~})$ \\
		
		\cline{2-4}
		
		& \rotatebox[origin=c]{90}{~S1.2~} &
		$(\text{-2~}, \text{~1~})$ &
		$(\text{~1~}, \text{~0~})$ \\
		
		\hline
	\end{tabular}
	
	\begin{tikzpicture}[remember picture,overlay]
		\draw[rounded corners,cyan!85!black,thick]
		([shift={(-0.5\tabcolsep,-1.0ex)}]pic cs:startup)
		rectangle
		([shift={(0.5\tabcolsep,2.5ex)}]pic cs:endup);
	\end{tikzpicture}
\end{table}

In \tabref{tab:decision_making}, the determination of the Nash equilibrium for the previously merged payoff matrix $\mathcal{D}$ is shown. For this purpose, the strategy combination with the highest payoffs for both cooperation candidates is selected. If there is only one Nash equilibrium (condition $1$) that corresponds to the original cooperation recommendation (condition $2$), a clear, joint solution has been found. If condition $1$ does not apply, the cooperation is rejected for safety reasons.

\subsubsection{Execution of Cooperative Maneuvers}
\label{subsubsec:execution_of_cooperative_maneuvers}

Within our approach, the determination of a Nash equilibrium is defined as signing a binding contract for performing the agreed strategy combination. Following an update of the situational overview, the planned cooperative maneuver is executed using in-vehicle trajectory planning and control algorithms.

In order to achieve a higher safety level during execution, time bounds or reserved lane areas, for example, could be agreed in advance. Further cooperative maneuvers may be recommended by the \TMS{} until the participating \CHAVs{} leave the intersection area.

%% file: sections/sect_evaluation.tex
\section{\uppercase{Evaluation}}
\label{sec:evaluation}

\noindent The evaluation of our cooperation approach builds upon a comprehensive simulation setup that includes virtual sensors, controllers for the \TMS{} and all vehicles involved, local situational overviews and internal states for all automated participants, specialized subscription and unsubscription procedures, a limited, prototypical implementation of the proposed approach as well as an omniscient evaluation framework with extensive monitoring and logging capabilities.

Despite the large implementation effort, we are aware that our setup is useful only for demonstration purposes, but not sufficient to fully answer the second and third research questions. We therefore plan to enhance this setup in follow-up activities and to publish more differentiated results in further contributions.

\subsection{Implementation}
\label{subsec:implementation}

To evaluate the cooperation approach simulatively, the developed concepts were prototypically implemented as a Python wrapper for the simulator \emph{Simulation of Urban Mobility} (\SUMO{}, version 1.1.0). As a widely used and commonly accepted open source traffic simulation, \SUMO{} offers a large community, extensive documentation, numerous application examples, and a plethora of scientific publications.

The simulation setup employs a TCP-based architecture and makes use of \emph{Traffic Control Interface} (\TraCI{}) to manipulate the behavior of all automated traffic participants. As a first step, a \emph{\TMS{} controller} is instantiated into the model. In each simulation step, an external simulation routine is triggered first, which adds new road users to the simulation. Each newly created vehicle is then equipped with an external \emph{\CHAV{}} or \emph{\HV{} controller} that can influence the participant's behavior in all subsequent steps.

As a result of the controller initialization, selected parameters of the \SUMO{} default vehicle models are overwritten with suitable values to represent the typical characteristics of \CHAVs{} and \HVs{} described in \secref{sec:top-level_architecture}, which are compared in \tabref{tab:comparison_of_vehicle_controller_parameters}.

\begin{table}[h]
	%\vspace{-0.2cm}
	\caption{Comparison of Vehicle Controller Parameters}
	\label{tab:comparison_of_vehicle_controller_parameters}
	\centering
	\begin{tabular}{|c|c|c|}
		\hline
		
		Parameter &
		\CHAVC{} &
		\HVC{} \\
		
		\hline
		
		Maximum Speed $v_{max}$ &
		$50$\,km/h &
		$60$\,km/h \\
		
		%Speed Factor $f_v$ &
		%$1$ &
		%(default) \\
		
		Speed Deviation $\delta_v$ &
		$0$ &
		$0.1$ \\
		
		Driver Imperfection $\sigma$ &
		$0.1$ &
		$0.5$ \\
		
		Reaction Time $t_r$ &
		$0.6$\,s &
		$1$\,s \\
		
		\hline
	\end{tabular}
\end{table}

As they approach the intersection, all \CHAVs{} subscribe themselves by calling respective methods of the \TMS{} controller. The \TMS{} senses the traffic situation at the intersection using virtual sensors and generates simplified cooperation recommendations for \CLCs{}, which are then forwarded to the \CHAV{} controllers. In order to evaluate the strategy combinations independently, each vehicle uses both virtual sensors and a randomly generated cooperative factor $f_c \in [0,1]$ that models the general willingness to cooperate. Following the evaluations, the two payoff matrices are exchanged via further method calls and a decision is made. If the \CHAVs{} decide to execute the maneuver, the driving behavior of the involved vehicles is manipulated by calling the \texttt{changeLane(\dots)} and \texttt{slowDown(\dots)} \TraCI{} methods. If not, the vehicles continue their journey without being affected.

\subsection{Execution}
\label{subsec:evaluation}

A simulation model of the research intersection in Brunswick provided by the \emph{Institute of Transportation Systems} of the \emph{German Aerospace Center} serves as a basis for the simulation-based evaluation. It comprises a true-to-scale representation of the physical intersection as part of the \emph{Application Platform for Intelligent Mobility} including an integrated traffic light sequence and a realistic lane layout. Besides five incoming lanes from east, south, and west as well as three incoming lanes from north, each intersection arm has two additional outgoing lanes.

The simulation model also includes recorded, real traffic data with a duration of one hour, in which arrival times and lanes, types, and destinations of all traffic participants (passenger cars, trucks, cyclists, and pedestrians) are defined. A previously created, deterministic mechanism divides passenger cars into \CHAVs{} and \HVs{} according to a selected distribution. In preparation for the simulation experiments, we also instantiated two vehicles with a fixed position and a constant speed of $0$\,km/h used as \emph{static obstacles} to provoke \CLCs{} in the western and southern approach in close proximity to the intersection area.

The first component of the evaluation was the execution of eleven simulation runs with an increasing percentage of \CHAVs{} ($0$, $10$, \dots, $100$\,\%) without being influenced by our cooperation approach. The goal of these simulations was to create a baseline for assessing the impact of cooperation. The trucks included in the traffic data were exclusively used without any automated controller. In addition, the behavior of cyclists and pedestrians remained unchanged. To increase the variance, ten repetitions of each simulation run were performed with different allocations of \CHAV{} and \HV{} controllers. The generated simulation results thus contain ten hours of simulated traffic per run, in each of which approximately $19.200$ \CHAVs{}, \HVs{}, and trucks as well as $1.200$ cyclists and $1.860$ pedestrians pass the intersection area.

As a second evaluation component, eleven additional simulation runs with the same allocations of \CHAV{} and \HV{} controllers were carried out, incorporating the presented cooperation approach. By using the same simulation environment and the same input data, the simulation results can be used to assess the impact of cooperation on traffic efficiency.

\subsection{Results}
\label{subsec:results}

Since the first research question has already been addressed by the presentation of our cooperation approach in \secref{sec:cooperation_approach}, we will focus on our simulation results to give some first hints on answering the subsequent questions. At first, a look into the global observations shall give an impression on the effects of automation and cooperation in general. A second rather detailed look focuses on individual parts of the intersection and investigates the influence of cooperative maneuvers on individual road users.

The potential options for assessing traffic efficiency are manifold. While municipal administrations usually consider capacities or maximum congestion lengths, human drivers or passengers evaluate traffic efficiency on the basis of crossing durations or the presence of directly perceptible delays. As the evaluation is based on a fixed simulation environment and real traffic data, it does not make sense to consider the junction's capacity as no vehicle gets stuck. Instead, the crossing duration will be taken into account, as it implicitly includes time losses and waiting times.

\subsubsection{Global Observations}
\label{subsubsec:global_observations}

\begin{figure}
	\centering
	\begin{tikzpicture}
		\begin{axis}[width=0.82\linewidth,
			scale only axis,
			xmajorgrids=true,
			xminorgrids=true,
			ymajorgrids=true,
			minor x tick num=1,
			enlarge x limits=0.05,
			enlarge y limits=0.05,
			grid style={line width=.1pt, draw=gray!10},
			major grid style={line width=.2pt,draw=gray!15},
			legend cell align={left},
			legend pos=north west,
			xlabel near ticks,
			ylabel near ticks,
			ytick={0,100,200,300,400,500},
			xlabel={\footnotesize Percentage of \CHAVs{} [\%]},
			ylabel={\footnotesize Number of Cooperating Vehicles [\#]}]
			
			\pgfplotsset{every tick label/.append style={font=\scriptsize}}
			
			\addplot[cyan!85!black,thick,mark size=1.5,mark=*] table{results/01_number_of_cooperating_vehicles_total.dat};
			\addlegendentry{\footnotesize Total}
			
			\addplot[cyan!75!black,thick,mark size=1.5,mark=square*] table{results/01_number_of_cooperating_vehicles_west.dat};
			\addlegendentry{\footnotesize West}
			
			\addplot[cyan!65!black,thick,mark size=1.5,mark=triangle*] table{results/01_number_of_cooperating_vehicles_south.dat};
			\addlegendentry{\footnotesize South}
			
			\addplot[cyan!55!black,thick,mark size=1.5,mark=x] table{results/01_number_of_cooperating_vehicles_north.dat};
			\addlegendentry{\footnotesize North}
		\end{axis}
	\end{tikzpicture}
	\caption{Numbers of Cooperating Vehicles}
	\label{fig:numbers_of_cooperating_vehicles}
\end{figure}
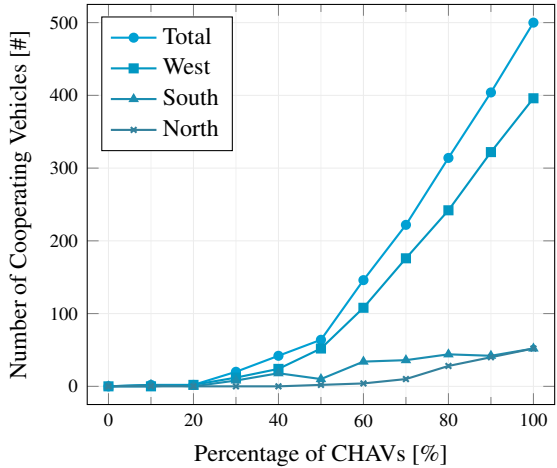

Starting with \figref{fig:numbers_of_cooperating_vehicles}, it can be observed that the total number of cooperating vehicles grows with an increasing percentage of \CHAVs{} from $2$ at a rate of $10$\,\% to $500$ at a rate of $100$\,\%. A division of the number of vehicles by two results in the number of cooperative maneuvers, since one cooperation always involves two participating \CHAVs{}. As expected, a majority of cooperative maneuvers can be observed on the lanes coming from west and south.

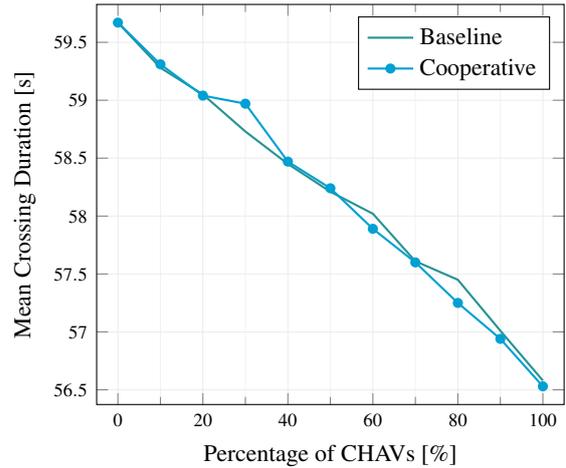
\begin{figure}
	\centering
	\begin{tikzpicture}
		\begin{axis}[width=0.82\linewidth,
			scale only axis,
			xmajorgrids=true,
			xminorgrids=true,
			ymajorgrids=true,
			minor x tick num=1,
			enlarge x limits=0.05,
			enlarge y limits=0.05,
			grid style={line width=.1pt, draw=gray!10},
			major grid style={line width=.2pt,draw=gray!15},
			legend cell align={left},
			legend pos=north east,
			xlabel near ticks,
			ylabel near ticks,
			ytick={56.5,57,57.5,58,58.5,59,59.5},
			xlabel={\footnotesize Percentage of \CHAVs{} [\%]},
			ylabel={\footnotesize Mean Crossing Duration [s]}]
			
			\pgfplotsset{every tick label/.append style={font=\scriptsize}}
			
			\addplot[teal!85!white,thick] table{results/02_mean_crossing_duration_base.dat};
			\addlegendentry{\footnotesize Baseline}
			
			\addplot[cyan!85!black,thick,mark size=1.5,mark=*] table{results/02_mean_crossing_duration_coop.dat};
			\addlegendentry{\footnotesize Cooperative}
		\end{axis}
	\end{tikzpicture}
	\caption{Mean Crossing Durations}
	\label{fig:mean_crossing_durations}
\end{figure}

By putting this figure in direct relation with \figref{fig:mean_crossing_durations}, it can be determined that along with an increased percentage of \CHAVs{} -- with and without cooperation -- the mean crossing duration decreases, which can be expected through strict compliance with legal requirements such as minimum distance and maximum speed of \CHAVs{} in contrast to \HVs{}. It is noteworthy that only from a rate above $50$\,\% of \CHAVs{} cooperative maneuvers bring an additional, albeit small, improvement in comparison to the baseline without cooperation. At rates below they can even have a negative impact, \eg{} at $30$\,\% with a prolongation of the mean crossing duration by $0.24$\,s.

However, when interpreting the results it must be taken into account that with respect to the small size of the traffic area under consideration only minor time differences are to be expected, especially if the percentage of \CHAVs{} is low. A further limitation is the simplified implementation of the \TMS{} controller, which only generates cooperation recommendations for a selected set of traffic situations. As already mentioned in \secref{sec:state_of_the_art}, a number of publications exclusively deal with algorithms for possible path planning that could be applied to improve the quality of the generated recommendations in future work.

\subsubsection{Local Observations}
\label{subsubsec:local_observations}

In the following, we focus on the western intersection approach, where the largest number of cooperative maneuvers can be observed. Thereby, the influence of cooperative maneuvers on both cooperating and non-cooperating vehicles is investigated more deeply.

\begin{figure}
	\centering
	\begin{tikzpicture}
		\begin{axis}[width=0.82\linewidth,
			scale only axis,
			xmajorgrids=true,
			xminorgrids=true,
			ymajorgrids=true,
			yminorgrids=true,
			minor x tick num=1,
			minor y tick num=1,
			enlarge x limits=0.05,
			enlarge y limits=0.05,
			grid style={line width=.1pt, draw=gray!10},
			major grid style={line width=.2pt,draw=gray!15},
			legend cell align={left},
			legend pos=south east,
			boxplot/draw direction=y,
			xtick={1,3,5,7,9,11},
			xticklabels={$0$,$20$,$40$,$60$,$80$,$100$},
			ytick={-100,-75,-50,-25,0,25,50,75,100},
			yticklabels={,,$-50$,$-25$,$0$,$25$,$50$,,},
			xlabel near ticks,
			ylabel near ticks,
			xlabel={\footnotesize Percentage of \CHAVs{} [\%]},
			ylabel={\footnotesize Variation of Crossing Durations [s]}]
			
			\pgfplotsset{every tick label/.append style={font=\scriptsize}}
			
			% Cooperating Vehicles (0 percent)
			\addplot[boxplot={average=auto},cyan!85!black,fill=cyan!85!black,fill opacity=0.1] table[x index=0,y index=0,row sep=newline]{results/06_effects_on_cooperating_vehicles_0pct_west.dat};
			
			% Cooperating Vehicles (10 percent)
			\addplot[boxplot={average=auto},cyan!85!black,fill=cyan!85!black,fill opacity=0.1] table[x index=0,y index=0,row sep=newline]{results/06_effects_on_cooperating_vehicles_10pct_west.dat};
			
			% Cooperating Vehicles (20 percent)
			\addplot[boxplot={average=auto},cyan!85!black,fill=cyan!85!black,fill opacity=0.1] table[x index=0,y index=0,row sep=newline]{results/06_effects_on_cooperating_vehicles_20pct_west.dat};
			
			% Cooperating Vehicles (30 percent)
			\addplot[boxplot={average=auto},cyan!85!black,fill=cyan!85!black,fill opacity=0.1] table[x index=0,y index=0,row sep=newline]{results/06_effects_on_cooperating_vehicles_30pct_west.dat};
			
			% Cooperating Vehicles (40 percent)
			\addplot[boxplot={average=auto},cyan!85!black,fill=cyan!85!black,fill opacity=0.1] table[x index=0,y index=0,row sep=newline]{results/06_effects_on_cooperating_vehicles_40pct_west.dat};
			
			% Cooperating Vehicles (50 percent)
			\addplot[boxplot={average=auto},cyan!85!black,fill=cyan!85!black,fill opacity=0.1] table[x index=0,y index=0,row sep=newline]{results/06_effects_on_cooperating_vehicles_50pct_west.dat};
			
			% Cooperating Vehicles (60 percent)
			\addplot[boxplot={average=auto},cyan!85!black,fill=cyan!85!black,fill opacity=0.1] table[x index=0,y index=0,row sep=newline]{results/06_effects_on_cooperating_vehicles_60pct_west.dat};
			
			% Cooperating Vehicles (70 percent)
			\addplot[boxplot={average=auto},cyan!85!black,fill=cyan!85!black,fill opacity=0.1] table[x index=0,y index=0,row sep=newline]{results/06_effects_on_cooperating_vehicles_70pct_west.dat};
			
			% Cooperating Vehicles (80 percent)
			\addplot[boxplot={average=auto},cyan!85!black,fill=cyan!85!black,fill opacity=0.1] table[x index=0,y index=0,row sep=newline]{results/06_effects_on_cooperating_vehicles_80pct_west.dat};
			
			% Cooperating Vehicles (90 percent)
			\addplot[boxplot={average=auto},cyan!85!black,fill=cyan!85!black,fill opacity=0.1] table[x index=0,y index=0,row sep=newline]{results/06_effects_on_cooperating_vehicles_90pct_west.dat};
			
			% Cooperating Vehicles (100 percent)
			\addplot[boxplot={average=auto},cyan!85!black,fill=cyan!85!black,fill opacity=0.1] table[x index=0,y index=0,row sep=newline]{results/06_effects_on_cooperating_vehicles_100pct_west.dat};
		\end{axis}
	\end{tikzpicture}
	\caption{Effects on Cooperating Vehicles (West)}
	\label{fig:effects_on_cooperating_vehicles_west}
\end{figure}
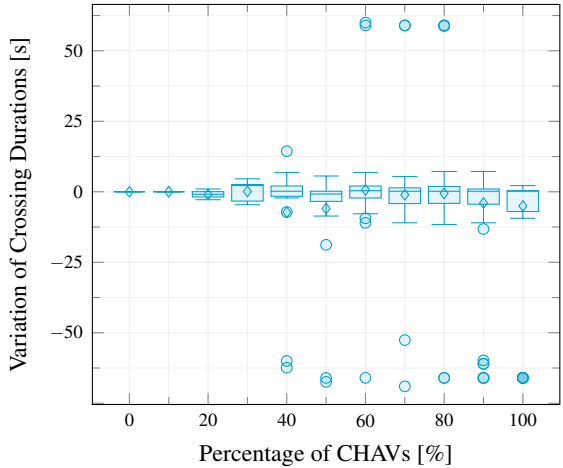

In \figsref{fig:effects_on_cooperating_vehicles_west}{fig:effects_on_non-cooperating_vehicles_west}, the variation of crossing durations is plotted on the y-axis along with the prevalence of \CHAV{} in percentage on the x-axis. A negative value represents an improvement (faster crossing) whereas a positive value represents a deterioration (slower crossing). The variation is to be interpreted in relation to the crossing time needed without cooperative maneuvers (baseline at constant $0$) and therefore only those vehicles were considered, which experienced a change of their crossing duration.

The effects on cooperating vehicles are depicted in \figref{fig:effects_on_cooperating_vehicles_west}. It can be seen that no cooperations were observed below a prevalence rate of \CHAVs{} of $20$\,\%. The medians of all subsequent distributions are always around zero and vary slightly in the first decimal place. The boxes which represent $50$\,\% of the vehicles are also around zero. However, with an increasing percentage of automated vehicles (exception at $60$\,\%) it can be stated that the vehicles from the upper quartile experience only a small time loss, whereas the vehicles from the lower quartile can achieve a greater time gain in comparison. The outliers to be observed are caused by the use of time-controlled traffic lights and the changed arrival times of the involved \CHAVs{}. As a red phase lasts up to $50$\,s, corresponding time gains and losses can be found. Since there are more outliers downwards than upwards, the mean values -- represented by diamonds -- are below zero above $50$\,\% of \CHAVs{} with an exception at $60$\,\% with $0.67$\,s.

\begin{figure}
	\centering
	\begin{tikzpicture}
		\begin{axis}[width=0.82\linewidth,
			scale only axis,
			xmajorgrids=true,
			xminorgrids=true,
			ymajorgrids=true,
			yminorgrids=true,
			minor x tick num=1,
			minor y tick num=1,
			enlarge x limits=0.05,
			enlarge y limits=0.05,
			grid style={line width=.1pt, draw=gray!10},
			major grid style={line width=.2pt,draw=gray!15},
			legend cell align={left},
			legend pos=south east,
			boxplot/draw direction=y,
			xtick={1,3,5,7,9,11},
			xticklabels={$0$,$20$,$40$,$60$,$80$,$100$},
			ytick={-100,-75,-50,-25,0,25,50,75,100},
			yticklabels={,$-75$,$-50$,$-25$,$0$,$25$,$50$,$75$,},
			xlabel near ticks,
			ylabel near ticks,
			xlabel={\footnotesize Percentage of \CHAVs{} [\%]},
			ylabel={\footnotesize Variation of Crossing Durations [s]}]
			
			\pgfplotsset{every tick label/.append style={font=\scriptsize}}
			
			% Non-Cooperating Vehicles (0 percent)
			\addplot[boxplot={average=auto},cyan!40!black,fill=cyan!40!black,fill opacity=0.1] table[x index=0,y index=0,row sep=newline]{results/06_effects_on_non-cooperating_vehicles_0pct_west.dat};
			
			% Non-Cooperating Vehicles (10 percent)
			\addplot[boxplot={average=auto},cyan!40!black,fill=cyan!40!black,fill opacity=0.1] table[x index=0,y index=0,row sep=newline]{results/06_effects_on_non-cooperating_vehicles_10pct_west.dat};
			
			% Non-Cooperating Vehicles (20 percent)
			\addplot[boxplot={average=auto},cyan!40!black,fill=cyan!40!black,fill opacity=0.1] table[x index=0,y index=0,row sep=newline]{results/06_effects_on_non-cooperating_vehicles_20pct_west.dat};
			
			% Non-Cooperating Vehicles (30 percent)
			\addplot[boxplot={average=auto},cyan!40!black,fill=cyan!40!black,fill opacity=0.1] table[x index=0,y index=0,row sep=newline]{results/06_effects_on_non-cooperating_vehicles_30pct_west.dat};
			
			% Non-Cooperating Vehicles (40 percent)
			\addplot[boxplot={average=auto},cyan!40!black,fill=cyan!40!black,fill opacity=0.1] table[x index=0,y index=0,row sep=newline]{results/06_effects_on_non-cooperating_vehicles_40pct_west.dat};
			
			% Non-Cooperating Vehicles (50 percent)
			\addplot[boxplot={average=auto},cyan!40!black,fill=cyan!40!black,fill opacity=0.1] table[x index=0,y index=0,row sep=newline]{results/06_effects_on_non-cooperating_vehicles_50pct_west.dat};
			
			% Non-Cooperating Vehicles (60 percent)
			\addplot[boxplot={average=auto},cyan!40!black,fill=cyan!40!black,fill opacity=0.1] table[x index=0,y index=0,row sep=newline]{results/06_effects_on_non-cooperating_vehicles_60pct_west.dat};
			
			% Non-Cooperating Vehicles (70 percent)
			\addplot[boxplot={average=auto},cyan!40!black,fill=cyan!40!black,fill opacity=0.1] table[x index=0,y index=0,row sep=newline]{results/06_effects_on_non-cooperating_vehicles_70pct_west.dat};
			
			% Non-Cooperating Vehicles (80 percent)
			\addplot[boxplot={average=auto},cyan!40!black,fill=cyan!40!black,fill opacity=0.1] table[x index=0,y index=0,row sep=newline]{results/06_effects_on_non-cooperating_vehicles_80pct_west.dat};
			
			% Non-Cooperating Vehicles (90 percent)
			\addplot[boxplot={average=auto},cyan!40!black,fill=cyan!40!black,fill opacity=0.1] table[x index=0,y index=0,row sep=newline]{results/06_effects_on_non-cooperating_vehicles_90pct_west.dat};
			
			% Non-Cooperating Vehicles (100 percent)
			\addplot[boxplot={average=auto},cyan!40!black,fill=cyan!40!black,fill opacity=0.1] table[x index=0,y index=0,row sep=newline]{results/06_effects_on_non-cooperating_vehicles_100pct_west.dat};
		\end{axis}
	\end{tikzpicture}
	\caption{Effects on Non-Cooperating Vehicles (West)}
	\label{fig:effects_on_non-cooperating_vehicles_west}
\end{figure}
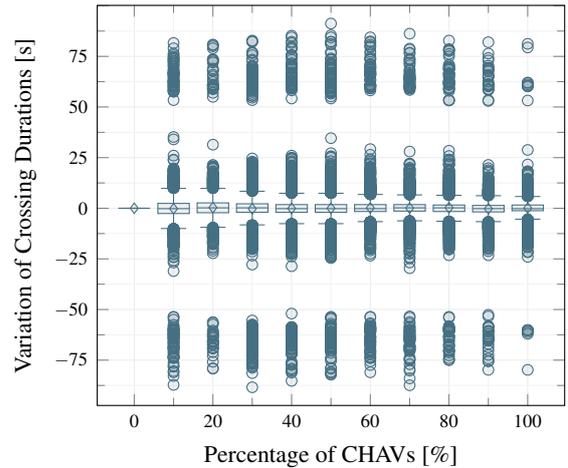

\figref{fig:effects_on_non-cooperating_vehicles_west} shows the effects on non-cooperating vehicles. The medians as well as the mean values and boxes are around zero, again. With an increasing percentage of \CHAVs{} the box sizes decrease slightly, probably due to a smoother traffic flow. It is noticeable that the only cooperation that is observed at $10$\,\% -- located in the southern approach -- has a clearly visible effect on the traffic situation in the western intersection arm. The reason for this seems to come from \SUMO{}, which varies some parameters of newly instantiated vehicles depending on the prevailing traffic situation. The distribution of outliers in terms of time gains and losses is again influenced by the behavior of the traffic lights, but can be regarded as balanced, so that neither a particular advantage nor disadvantage can be determined on non-cooperating vehicles.

%% file: sections/sect_conclusion.tex
\section{\uppercase{Conclusion}}
\label{sec:conclusion}

\noindent  In this paper, we proposed a decentralized, game-theoretic approach for the negotiation of cooperative maneuvers of \CHAVs{} at urban intersections. To support these maneuvers, we assumed the existence of an intelligent \TMS{} with a global view on the whole scenery in order to derive cooperation recommendations without directly controlling the \CHAVs{}. This combination enables a mutually consistent, distributed decision making on cooperative driving maneuvers, taking into account global and local goals. Due to the large implementation effort, the evaluation results are preliminary with regard to a possible increase in traffic efficiency and the identification of causal relationships. We therefore plan to enhance our evaluation, \eg{} by investigating the influence of different traffic densities or controller configurations, and to report on more elaborated results.

Until now, it can be stated that both cooperating and non-cooperating vehicles are influenced positively and negatively in the sense of shorter and longer crossing durations. A trend shows that the advantages for cooperating vehicles seem to be much greater than the disadvantages, although (so far) not for all percentages of \CHAVs{}. Positive and negative impacts on non-cooperating vehicles compensate each other. The results may vary depending on the geometry and other characteristics such as the number of the lanes.

Taking into account the fact that the approach only considers a single intersection without any cascading effects and that initially only two \CHAVs{} are able to perform a cooperative maneuver, it was possible to create a solid basis for further evaluations. The prototypical implementation of required methods for the generation of cooperation recommendations can certainly be improved on the basis of existing research. Also the \CHAV{} controller could be refined, \eg{} by including more realistic virtual sensors or behavioral predictions. In addition, we plan to extend our approach in a granted follow-up project to support the interaction of cooperating vehicle groups in significantly larger traffic areas with multiple intersections. The associated potential to increase traffic efficiency still offers plenty of scope for further research.